\documentclass{CSML}

\pdfoutput=1

\usepackage{lastpage}
\def\dOi{13(4:17)2017}
\lmcsheading%
{\dOi}
{1--\pageref{LastPage}}
{}
{}
{Feb.~22, 2016}
{Nov.~28, 2017}
{}


\usepackage[hidelinks]{hyperref}

\usepackage{url}
\usepackage{amssymb}
\usepackage{amsmath}
\usepackage{amsfonts}
\usepackage{mathrsfs}
\usepackage{stackrel}

\usepackage{ourmacros}

\begin{document}

\title[Branching Bisimilarity of Normed BPA]{Branching Bisimilarity of
Normed BPA Processes\\as a Rational Monoid}

\author[P.~Jan\v{c}ar]{Petr Jan\v{c}ar}
\address{Technical University Ostrava (FEI V\v{S}B-TUO), Czech Rep.}
\email{petr.jancar@vsb.cz}
\thanks{The work was supported 
by the Grant Agency of the Czech Rep., project GA\v{C}R:15-13784S}

\begin{abstract}
\noindent
The paper presents an elaborated and simplified version
	of the structural result 
for branching bisimilarity on normed BPA (Basic Process Algebra)
processes that was the crux of a conference paper by 
	Czerwi\'nski and Jan\v{c}ar (arxiv 7/2014 and LiCS 2015).
That paper focused on the computational complexity, 
and a NEXPTIME-upper bound has been derived; 
the authors built on the ideas by  Fu (ICALP 2013), and  
 strengthened his
decidability result. Later 
He and Huang announced the EXPTIME-completeness of this problem
(arxiv 1/2015, and LiCS 2015), giving a technical proof
for the EXPTIME membership. He and Huang indirectly 
acknowledge the decomposition ideas 
by Czerwi\'nski and Jan\v{c}ar	
on which they also built, 
	but it is difficult 
	to separate 
	their starting point  from 
their new ideas. 

One aim here is to present the previous decomposition result 
of Czerwi\'nski and Jan\v{c}ar	
	in a technically new framework,
noting that branching bisimulation equivalence
on normed BPA
	processes corresponds to a rational monoid (in the sense of
	[Sakarovitch, 1987]); in
particular it is shown that the mentioned
equivalence can be decided by normal-form computing deterministic
finite transducers. Another aim is to provide a complete description,
including an informal overview, that should also make clear how Fu's ideas 
were used, and to give 
all proofs in a form that should be readable and easily verifiable.
\end{abstract}

\maketitle

\section{Introduction}\label{sec:intro}

Bisimulation equivalence (or bisimilarity) is
a fundamental notion in theory of processes,
and the respective decidability and complexity questions are
a natural research topic; we can refer to~\cite{Srba:Roadmap:04} 
for an (updated) overview of the results in a specific
area of process rewrite systems.

One basic model of infinite-state systems is called Basic Process
Algebra (BPA), which can be naturally
related to context-free grammars in Greibach normal form.
Here the \emph{processes}
are identified with finite sequences of variables
(nonterminals); a process $A\alpha$ can change by performing an 
\emph{action}, denoted by $A\alpha\trans{a}\beta\alpha$, 
in which case its
leftmost variable $A$ is rewritten according to a grammar rule
$A\trans{}a\beta$ (presented rather as $A\trans{a}\beta$ in our context).

A seminal paper by Baeten, Bergstra and
Klop~\cite{DBLP:journals/jacm/BaetenBK93} showed the decidability of
bisimilarity on nBPA, i.e. on the \emph{normed} BPA processes,
where each variable can be stepwise
rewritten to the empty word;
this decidability result was later extended 
to the whole class BPA~\cite{DBLP:journals/iandc/ChristensenHS95}.
Further exploration placed the problem on nBPA even
in $\PTIME$~\cite{DBLP:journals/tcs/HirshfeldJM96}
(this problem is thus
$\PTIME$-complete~\cite{DBLP:journals/fac/BalcazarGS92});
we can refer to~\cite{CzerwinskiThesis} 
for further references and a way towards the so far best known upper bound.
The bisimilarity problem for the whole class BPA is known to be
$\EXPTIME$-hard~\cite{DBLP:journals/ipl/Kiefer13} and to belong to
 $2$-$\EXPTIME$
(claimed in~\cite{DBLP:conf/mfcs/BurkartCS95} and explicitly proven 
in~\cite{DBLP:journals/corr/abs-1207-2479}).

When also internal (unobservable) actions of systems are taken into
account, the most studied generalization of bisimilarity is \emph{weak
bisimilarity}~\cite{Milner:book} but the relevance of the finer
equivalence called~\emph{branching bisimilarity} is also
well argued~\cite{DBLP:journals/jacm/GlabbeekW96}. 

The (un)decidability status of weak bisimilarity on BPA, as well as on nBPA,
is still open, but we have  
the $\EXPTIME$-hardness result 
by Mayr~\cite{DBLP:journals/tcs/Mayr05} for weak bisimilarity
on nBPA. 
Similarly, the decidability status of weak bisimilarity is still open
in the case of (normed) Basic Parallel Processes, 
which is the parallel (or commutative) version of BPA. 

The situation seems more
favourable
in the case of branching bisimilarity.
It was first shown 
decidable for the normed 
Basic Parallel Processes~\cite{DBLP:journals/mst/CzerwinskiHL14}, and 
then Fu~\cite{DBLP:conf/icalp/Fu13} showed the 
decidability on nBPA.
A later paper~\cite{DBLP:conf/icalp/YinFHHT14} 
shows that the mentioned decidability results for branching
bisimilarity cannot be essentially extended, possibly with the
exception of the full 
classes of BPA processes and
of Basic Parallel Processes for which the decidability question remains
open.

The case of \emph{branching bisimilarity on nBPA} is the main topic of this
paper. We first note that
Fu's decidability result~\cite{DBLP:conf/icalp/Fu13}
 is substantially
 stronger than the previous results dealing with so called
 \emph{totally}
normed
BPA~\cite{DBLP:conf/cav/Huttel91,DBLP:journals/iandc/CaucalHT95}
(where no variable can ``disappear'' by unobservable actions).
In the case of totally normed BPA processes even
a polynomial time algorithm is suggested
in~\cite{DBLP:journals/corr/He14a}, building on the unique-decomposition results and techniques that were previously 
used in the case of (strong) bisimilarity on nBPA.

A crucial novel idea in Fu's decidability proof is a use of
the notion that can be called the \emph{class-change norm} 
(called the
branching norm in~\cite{DBLP:conf/icalp/Fu13});
while the standard norm counts
all steps in rewriting a process to the empty word,
the class-change norm only counts the steps 
that change the current equivalence-class. 
It is not clear how to
compute this norm directly but equivalent processes $\alpha\sim\beta$
must have the same class-change norm.
Another useful fact shown by Fu is that the relation
of $\alpha\gamma$ and $\beta\gamma$ (either
$\alpha\gamma\sim\beta\gamma$ or $\alpha\gamma\not\sim\beta\gamma$) is
determined solely 
by the \emph{redundant variables} w.r.t. $\gamma$, i.e. by those $X$
for which $X\gamma\sim \gamma$, independently of the string $\gamma$
itself.

This paper is based on the research reported on
in~\cite{DBLP:conf/lics/CzerwinskiJ15}, performed
with W. Czerwi\'{n}ski (see Author's acknowledgements).
The main new idea there was to use 
the \emph{decompositions} of processes that are \emph{relative} to
a given set of (redundant) variables; the notion is also based on the
(semantic) class-change norm.
This structural result is here a bit reworked 
and presented in a technically new framework;
it is shown that the quotient of branching bisimulation equivalence
on nBPA is a rational monoid
(in the sense of
Sakarovitch~\cite{DBLP:journals/iandc/Sakarovitch87}).
In particular, for a given nBPA system $\calG$ branching bisimilarity
 can be decided by a (canonical) normal-form computing deterministic
finite-state transducer $\calT^\calG$; to each process $\alpha$ it
computes the normal form $\calT^\calG(\alpha)$, which is a unique
process in the equivalence-class $[\alpha]_\sim$, and we have
$\alpha\sim\beta$ iff  $\calT^\calG(\alpha)=\calT^\calG(\beta)$.
The size of  $\calT^\calG$
can be easily 
bounded by 
an exponential function of the size of $\calG$.

We will not show a direct construction of $\calT^\calG$,
but we will show a quickly verifiable consistency
condition for any given
transducer $\calT$ that guarantees 
$\calT(\alpha)=\calT(\beta) \Rightarrow \alpha\sim\beta$; moreover,
 $\calT^\calG$ will be shown to satisfy this consistency condition 
 w.r.t. 
$\calG$.
This immediately yields a nondeterministic exponential-time
algorithm deciding 
branching bisimilarity on nBPA:
given $\calG$, $\alpha,\beta$, guess a transducer $\calT$ of at most
exponential size (in the size of $\calG$), check that $\calT$ 
	is
consistent with $\calG$,
and verify that 
	$\calT(\alpha)=\calT(\beta)$.

The problem for which Fu~\cite{DBLP:conf/icalp/Fu13}
showed the decidability (by an involved proof in a tableau framework)
is thus placed in $\NEXPTIME$. 	
Regarding the question of a lower bound,
Fu~\cite{DBLP:conf/icalp/Fu13} noted that the problem is
$\EXPTIME$-hard, which was later confirmed by Huang and
Yin~\cite{HY2017}. (More details about this interesting point are
given in Section~\ref{sec:overview}.)
In Section~\ref{sec:complexity} we also add some remarks on a possible
construction of the canonical transducer $\calT^\calG$ in
deterministic exponential time; this seems to be (at least implicitly)
related to 
the paper by  He and Huang~\cite{DBLP:conf/lics/HeH15} that announced
\EXPTIME-completeness.

\textit{Structure of the paper.}
In Section~\ref{sec:prelim} we define the used notions and make some
simple observations. 
Section~\ref{sec:overview} gives an informal overview, which is then
formalized in Section~\ref{sec:proof}. 
Section~\ref{sec:overview} also contains a remark on the lower complexity
bound, and
Section~\ref{sec:complexity} adds some further remarks.

\section{Preliminaries}\label{sec:prelim}

We put $\N=\{0,1,2,\dots\}$,
and $[i,j]=\{i,i{+}1,\dots,j\}$  for $i,j\in\N$. 

For a set $M$, by $M^*$ we denote the set of finite sequences of
elements of $M$, also
called \emph{words}, or \emph{strings}, \emph{over} $M$; by $\varepsilon$
we denote the empty string.
For $\alpha\in M^*$, by $|\alpha|$ we denote
its length.

\subsection*{Labelled transition systems}

A \emph{labelled transition system}, an \emph{LTS} for short,
 is a tuple
$$\calL=(\calS,\act, (\trans{a})_{a\in\act})$$
where $\calS$ is the set of \emph{states},
 $\act$ is the set of \emph{actions} 
 and
 $\trans{a}\subseteq \calS\times\calS$ is the set of
 \emph{$a$-labelled transitions}.
 We reserve the symbol
 \begin{center} 
 $\tau$ for  the (unique) \emph{silent action};
the  \emph{visible actions} are the elements of $\act\smallsetminus\{\tau\}$.
\end{center}
We write $s\trans{a}t$ rather than $(s,t)\in\trans{a}$ (for $a\in\act$),
 and we define
 $s\trans{w}t$ for $w\in\act^*$ inductively:
 $s\trans{\eps}s$; if $s\trans{a}s'$ and
 $s'\trans{u}t$, then $s\trans{au}t$. By $s\trans{w}t$ we sometimes
also refer to a concrete respective path from $s$ to $t$ in $\calL$.
(We do not exclude cycles in the paths.)

\subsection*{Branching bisimilarity, i.e., branching bisimulation
equivalence $\sim$}

Given an LTS 
$\calL=(\calS,\act, (\trans{a})_{a\in\act})$,
a 
relation $\calB\subseteq \calS\times\calS$ is a \emph{branching
 bisimulation} in $\calL$ if 
 for any $(s,t)\in\calB$
 the following two conditions hold:
		\begin{enumerate}[label={\roman*)}]
\item
for any $a\in\act$, any  
\emph{move} $s\trans{a}s'$
		 \emph{can be matched from} $t$, i.e.,
\begin{enumerate}[label={\alph*)}]		 
	 \item		 
$a=\tau$ and $(s',t)\in \calB$, or
\item
 there is a path
 $t=t_0\trans{\tau}t_1\trans{\tau}\cdots\trans{\tau}t_k\trans{a}t'$
(for some $k\geq 0$)
such that $(s',t')\in\calB$ and
$(s,t_i)\in\calB$ for all $i\in [1,k]$;
\end{enumerate}
	 \item
for any $a\in\act$, any
		 \emph{move} $t\trans{a}t'$
		 \emph{can be matched from} $s$, i.e.,
\begin{enumerate}[label={\alph*)}]
	 \item		 
$a=\tau$ and $(s,t')\in \calB$, or
\item
 there is a path
 $s=s_0\trans{\tau}s_1\trans{\tau}\cdots\trans{\tau}s_k\trans{a}s'$
(for some $k\geq 0$) 
such that $(s',t')\in\calB$ and
$(s_i,t)\in\calB$ for all $i\in [1,k]$.
\end{enumerate}
\end{enumerate}
 By $s\sim t$, to be read as ``states $s,t$ are \emph{branching bisimilar}'',
 we denote that there is a branching bisimulation containing $(s,t)$.
We can easily verify the standard facts that
$\sim\,\subseteq\calS\times\calS$ is the union of
all branching bisimulations (in $\calL$),
and thus the largest branching bisimulation in $\calL$, and that $\sim$ 
is an equivalence
 relation.

\subsection*{Class-changing transitions, and class-change norm $\bnorm{s}$.}

Assuming an LTS $\calL=(\calS,\act, (\trans{a})_{a\in\act})$,
we now introduce a few notions and make simple observations
that turn out to be very useful for our aims.
We say that 
\begin{center}
	a \emph{transition}
$s\trans{a}s'$ 
	is 
\emph{class-changing} if $s\not\sim s'$.
\end{center}
Hence a class-changing transition leads from one equivalence class of
$\sim$ to a different class.
We note that $s\sim t$ and $s\trans{a}s'$ entails 
that either $a=\tau$ and  $s\trans{a}s'$ is not class-changing (in
which case $s'\sim t$), or there is a path
$t=t_0\trans{\tau}t_1\trans{\tau}\cdots\trans{\tau}t_k\trans{a}t'$
(for some $k\geq 0$) 
where no transition in the path
$t_0\trans{\tau}t_1\trans{\tau}\cdots \trans{\tau}t_{k}$ is
class-changing (hence $t=t_0\sim t_1\sim \cdots\sim t_k$) and  $s'\sim t'$; in the latter case, the transition
$t_k\trans{a}t'$ is class-changing iff
$s\trans{a}s'$ is class-changing. 

We introduce the class-change norm $\bnorm{s}$ as 
the ``class-change distance'' of $s$ to
 the silent states.
A \emph{state} $s$ is
\emph{silent} if $s\trans{w}s'$ entails $w\in\{\tau\}^*$.
(Hence we can never perform a visible action when starting from a
silent state.)
Let $S_{\sil}$ be the \emph{set of silent states}
(in the assumed LTS $\calL=(S,\act,(\trans{a})_{a\in\act})$); $S_{\sil}$ is 
obviously a (maybe empty) equivalence class of $\sim$ (since the set
$\{ (s,t)\mid s,t\in
S_{\sil}\}$ is a branching bisimulation, and $s\in S_{\sil}$, $t\not\in
S_{\sil}$ implies $s\not\sim t$).

By $\bnorm{s}$ we denote the \emph{class-change norm} of $s$,
the \emph{cc-norm} for short,
which is the least $\ell\in\N$ such that there is a path 
$s=s_{0}\trans{a_1}s_1\trans{a_2}\cdots \trans{a_k}s_k\in  S_{\sil}$ that
has precisely $\ell$
class-changing transitions;
 we put $\bnorm{s}=\omega$ if $S_{\sil}$ is not
reachable from $s$. Hence $\bnorm{s}=0$ iff $s\in S_{\sil}$.
The previous discussion (of matching $s\trans{a}s'$ from $t$ when
$s\sim t$) easily yields the following fact:

\begin{obs}\label{obs:ccnorm}
	If $s\sim t$, then $\bnorm{s}=\bnorm{t}$.
 \end{obs}

\begin{rem} 
The cc-norm 
was introduced by Fu in~\cite{DBLP:conf/icalp/Fu13},
who used the name ``branching norm'' and a slightly different
form; formally his norm also counts the visible transitions no matter
if they are class-changing or not but this is no crucial difference,
in fact. 
\end{rem}

\subsection*{BPA systems and processes.}

We view a~\emph{BPA system} (where BPA stands for Basic Process Algebra)
 as a context-free grammar in Greibach
 normal form, with no starting variable (nonterminal). We denote it as
$$\calG=(\var,\act,\rules)$$
where
 $\var$ is a finite set of \emph{variables} (or nonterminals), $\act$
 is a finite set of \emph{actions} (or terminals), which can contain
 the \emph{silent action} $\tau$, and $\rules$ is a finite set of
 \emph{rules} of the form
 $A\trans{a}\alpha$ where $A\in\var$, $a\in\act$,
 $\alpha\in\var^*$. 

A BPA system $\calG=(\var,\act,\rules)$
has the associated LTS 
$$\calL_\calG=(\var^*,\act,(\trans{a})_{a\in\act})$$
where 
each rule 
 $A\trans{a}\alpha$ in $\rules$ induces the transitions
$A\beta\trans{a}\alpha\beta$ for all $\beta\in\var^*$.
The states of $\calL_\calG$, i.e. the strings of variables,
are also called 
\emph{processes}.

\subsection*{Standard (syntactic) norm
$\norm{\alpha}$, and normed BPA systems (nBPA)}

Given a BPA system  $\calG=(\var,\act,\rules)$, 
the \emph{norm} $\norm{\alpha}$ of  
$\alpha\in\var^*$ 
is the length $|w|$ of a shortest
$w\in\act^*$ such that $\alpha\trans{w}\eps$; we put 
$\norm{\alpha}=\omega$ when there is no such $w$ (where $\omega$
stands for an ``infinite amount'').
We say that $\alpha$ is \emph{normed} if $\norm{\alpha}\in\N$
(i.e., if $\alpha\trans{w}\eps$ for some $w$).
The \emph{BPA system} $\calG$
is \emph{normed},
an  \emph{nBPA system} for short, 
if each variable 
$A\in\var$
is normed.

A \emph{transition} $\alpha\trans{a}\beta$ is \emph{norm-reducing} if
$\norm{\alpha}>\norm{\beta}$, in which case
$\norm{\beta}=\norm{\alpha}{-}1$,
in fact. If $\norm{\alpha}=\omega$, then there is no norm-reducing
transition  $\alpha\trans{a}\beta$.
The facts captured by the next proposition are standard; they
also entail that we can check in polynomial time whether a BPA system
is normed.

\begin{prop}\label{prop:standnorm}
Given a BPA system $\calG=(\var,\act,\rules)$, we have:
	\begin{enumerate}
		\item $\norm{\eps}=0$. 
		\item
 $\norm{\alpha\beta}= \norm{\alpha}+\norm{\beta}$ (assuming
	$\omega+z=z+\omega=\omega$ for any $z\in\N\cup\{\omega\}$).
\item
 $\norm{A}=1+\norm{\alpha}$ for a norm-reducing rule
	$A\trans{a}\alpha$, if $\norm{A}\in\N$.
\item
There is a polynomial-time algorithm  that  computes $\norm{A}$ for each
$A\in\var$.
\item
The finite
values $\norm{A}$ are at most exponential in the size of $\calG$.
	\end{enumerate}
	\end{prop}	
We note in particular
that the algorithm in the point (4) can naturally use dynamic
programming: We first temporarily assume
 $\norm{A}=\omega$ (the norm is infinite) for all variables; this
also temporarily yields
$\norm{\alpha}=\omega$ for all rhs (right-hand sides) of
the rules $A\trans{a}\alpha$, except of $\alpha=\eps$ since we put
$\norm{\eps}=0$ definitively.
Now we repeatedly look for a
variable $A$ with a temporary norm
that has a rule $A\trans{a}\alpha$ with the least definitive
$\norm{\alpha}\in\N$; for such $A$ we put 
$\norm{A}=1+\norm{\alpha}$ definitively (all variables in $\alpha$
have the definitive norms already),
and we recompute the temporary norms of
the right-hands sides of the rules in $\calR$ accordingly.
After this repeated process finishes, all temporary cases
$\norm{A}=\omega$ become also definitive.

\subsection*{Branching bisimilarity problem for nBPA}

The \emph{nBPA-bbis problem} asks,
given an nBPA system $\calG=(\var,\act,\rules)$ and two processes
	$\alpha,\beta\in\var^*$, if $\alpha\sim\beta$, i.e.,
	if $\alpha$ and $\beta$ are branching bisimilar as the states
	in $\calL_\calG$.

We add a remark on
 $\bnorm{\alpha}$, which refers to the (``semantic'') cc-norm of
	$\alpha$ in
	$\calL_{\calG}$.
	By Observation~\ref{obs:ccnorm} we know that $\alpha\sim\beta$
	implies $\bnorm{\alpha}=\bnorm{\beta}$. 
	We have shown how to compute the (``syntactic'') norm
	$\norm{\alpha}$, but it is unclear how to compute 
$\bnorm{\alpha}$. Nevertheless, since $\varepsilon$ is a silent state
in $\calL_\calG$, we can easily
	observe that $\bnorm{\alpha}\leq\norm{\alpha}$ 
	(and $\bnorm{A}$ is thus at most exponential by (5) in
	Prop.~\ref{prop:standnorm}).

\section{Informal overview}\label{sec:overview}

Here we sketch some informal ideas that are elaborated in
Section~\ref{sec:proof}.
We also use a small, but important, example; the example is inspired by a
recent work of Huang and Yin~\cite{HY2017}, which is further discussed
below in an additional remark on the lower complexity bound for
the nBPA-bbis problem.

Let us consider the following BPA system
$\calG=(\var,\act,\rules)$ where
\begin{itemize}
	\item
		$\var=\{A,B,C,E,F_B,[F_B,A]\}$ (hence $|\var|=6$, since
		$[F_B,A]$ is one symbol),
	\item
		$\act=\{\tau\}\cup\{a,b,c,e,f^1_B,f^2_B\}$,
	\item
$\rules=
\{A\trans{a}\varepsilon, B\trans{b}\varepsilon, 
		C\trans{c}\varepsilon, E\trans{e}\varepsilon,$
\\
		\hspace*{2.6em}$F_B\trans{\tau}\varepsilon,
		F_B\trans{f^1_B}\varepsilon,
		$
\\		
	\hspace*{2.6em}$A\trans{f^1_B}[F_B,A],
B\trans{f^1_B}B,
C\trans{f^1_B}E,$ 
\\
		\hspace*{2.6em}$[F_B,A]\trans{a}\varepsilon,
		[F_B,A]\trans{f^1_B}[F_B,A],$
\\
		\hspace*{2.6em}$[F_B,A]\trans{f^2_B}\eps,A\trans{f^2_B}F_B\}$.
\end{itemize}		
All variables are normed, we even have $\norm{X}=1$ for all $X\in\var$.

In the LTS $\calL_\calG$
we have, e.g., $F_BAABACCAB\sim AABACCAB$ but 
$F_BAACABCAB\not\sim AACABCAB$. More generally for any
$\alpha\in\{A,B,C\}^*$ we have 
\begin{equation}\label{eq:firstBexample}
	\textnormal{$F_B\alpha\sim\alpha$ iff
	$\alpha=\alpha'B\alpha''$ where $\alpha'\in\{A\}^*$.}
\end{equation}	
We can view $F_B$ as a ``claim'' that $\alpha$ satisfies ``first
$B$'', which means that $\alpha$ 
 contains $B$ as the first
(i.e., leftmost) occurrence of a symbol from $\{B,C\}$.
We leave the verification of~(\ref{eq:firstBexample}) 
as an interesting small exercise, since it is not crucial for us.

The example illustrates that a variable can be or not be
``redundant'' (w.r.t. $\sim$),
depending on the ``suffix''; formally we say that
$R_\gamma=\{X\in\var\mid X\gamma\sim\gamma\}$ is the set of
redundant variables {w.r.t.}\ $\gamma$.  
In the example, the
condition characterizing the strings $\alpha\in\{A,B,C\}^*$ for which $F_B\alpha\sim\alpha$
 is regular, i.e.
checkable by a finite automaton.
It turns out to be an important fact that
each nBPA system $\calG$ has an associated finite automaton
$\calF^\calG$ that determines 
the set $R_\gamma$
after reading $\gamma$.
Moreover, it turns out possible, and convenient, to let the automaton 
$\calF^\calG$
read its input $\gamma$ from right to left and use 
the respective sets $R\subseteq\var$ as its control states; the
automaton starts in the
initial state $R_\varepsilon$ (which might be the empty set) and after
reading $\gamma$ (from right to left) it enters the state $R_\gamma$.
Its transitions are thus of the form
$R_{A\gamma}\lltrans{A}{}R_\gamma$, in the notation that visualizes reading
from right to left. (In Section~\ref{sec:proof} we also show the
soundness:
$R_{\gamma}=R_\delta$ entails $R_{A\gamma}=R_{A\delta}$.)
These ideas were already
developed by Fu~\cite{DBLP:conf/icalp/Fu13}
(though he did not mention the automaton explicitly).

The above example can be generalized to show
that the automaton $\calF^\calG$ can have exponentially many states
$R_\gamma$ (w.r.t. the size of
the given nBPA system $\calG$):
we can add several other pairs $\{B',C'\}$ of variables, 
with the respective variables $F_{B'}, [F_{B'},..]$ and the respective actions and rules.
The issue of exponentially many sets of redundant variables
is dealt with in~\cite{HY2017} in more detail. 

 \medskip

\begin{rem} 
In~\cite{DBLP:conf/icalp/Fu13} there was also a note
 saying that the nBPA-bbis problem can be shown
\EXPTIME-hard by a slight modification of Mayr's proof for weak
bisimilarity~\cite{DBLP:journals/tcs/Mayr05}. 
Though this note was repeated in further works, no rigorous proof 
 was given
(as pointed out in the first version of this 
paper~\cite{DBLP:journals/corr/Jancar16}).
The mentioned ``slight modification'' has turned out to be not so
obvious, but the \EXPTIME-hardness has been recently rigorously confirmed by 
 Huang and Yin~\cite{HY2017}. 
 
Mayr's \EXPTIME-hardness 
proof~\cite{DBLP:journals/tcs/Mayr05} uses a reduction from the ALBA
problem (Alternating Linear Bounded Automata acceptance), a standard
$\EXPTIME$-complete problem.  
Huang and Yin~\cite{HY2017} decided to use the Hit-or-Run game 
for their reduction; this \EXPTIME-complete problem 
was used by Kiefer~\cite{DBLP:journals/ipl/Kiefer13} to show
the \EXPTIME-hardness of strong bisimilarity for (general) BPA
systems.
The \EXPTIME-hardness of the Hit-or-Run game was also
established by a reduction from the ALBA problem.
It is worth to note that it is also possible 
to modify Mayr's reduction~\cite{DBLP:journals/tcs/Mayr05} by the new idea 
of~\cite{HY2017}, to yield  
\EXPTIME-hardness of the nBPA-bbis problem by a direct
reduction from the ALBA problem;  the above example 
(inspired by~\cite{HY2017}) captures the essence since it shows how it
is possible to
``remember'' an ALBA configuration by the current set of redundant
variables. (If $\alpha$ is a sequence of ALBA configurations, 
then 
$R_\alpha$ determines the leftmost configuration in the sequence; we
use a special pair $\{[i,0],[i,1]\}$ of variables (like $\{B,C\}$
in the example), with the respective additional variables, actions and
rules, to ``remember'' if the $i$-th position is $0$ or $1$.)

Hence Fu's remark in~\cite{DBLP:conf/icalp/Fu13} can be viewed as
correct in the end, though it has not been straightforward to come with the
appropriate ``slight'' modification.  
\end{rem}
\medskip

The contribution of this paper captures the decomposition ideas 
from~\cite{DBLP:conf/lics/CzerwinskiJ15}.
The above discussed automaton $\calF^\calG$, satisfying
$R_{\alpha}\lltrans{\alpha}{}R_\varepsilon$, can be enhanced to become
 a
finite-state transducer $\calT^\calG$ (corresponding to a given nBPA system
$\calG=(\var,\act,\calR)$) that translates its input $\alpha$ into a string
$\calT^\calG(\alpha)$, processing $\alpha$ from right to left; this is denoted 
$R_{\alpha}\lltrans{\alpha}{\beta}R_\varepsilon$ where 
$\beta=\calT^\calG(\alpha)$.
For the uniqueness of the ``canonical transducer''
$\calT^\calG$ 
we use a linear order on
$\var$ and take $\calT^\calG(\alpha)$ as the lexicographically
smallest string 
among the longest redundancy-free strings from the
equivalence class $[\alpha]_\sim$. (Here the lexicographic
order of two different strings is determined by the first position \emph{from
the right} where the strings differ.)
 By the
redundancy-freeness of a string $\beta$ we mean that $\beta=\beta'A\beta''$
entails that $A\beta''\not\sim\beta''$ (i.e., $A\not\in R_{\beta''}$).
We recall that $\beta\in [\alpha]_\sim$ entails $\bnorm{\beta}=\bnorm{\alpha}$ (by
Observation~\ref{obs:ccnorm}), 
and we can observe that  
$|\beta|\leq\bnorm{\beta}$ when 
$\beta$ is redundancy-free
(since any path  $A\beta''\trans{u}\beta''$ contains at least one
class-changing transition when $A\beta''\not\sim \beta''$).

We will verify the soundness of the
above definition of the canonical transducer $\calT^\calG$
(for any \emph{normed} BPA system $\calG$). We thus get
\begin{center}
$\alpha\sim\beta$ (in $\calL_\calG$) iff 
$\calT^\calG(\alpha)=\calT^\calG(\beta)$.
\end{center}
	We also note the idempotency 
$\calT^\calG(\calT^\calG(\alpha))=\calT^\calG(\alpha)$, and the fact
that
$\calT^\calG(\alpha)$ can be naturally viewed as the \emph{normal
form} (or the
\emph{prime decomposition}) of $\alpha$; two strings $\alpha,\beta$ are
equivalent (meaning branching bisimilar) iff they have the same
normal forms (the same prime decompositions).
We also note that generally we do not have
$\calT^\calG(\alpha\gamma)=\calT^\calG(\alpha)\,\calT^\calG(\gamma)$,
since the decomposition is more subtle: 
we have 
$\calT^\calG(\alpha\gamma)=\calT^\calG_{R_\gamma}(\alpha)\,\calT^\calG_{R_\varepsilon}(\gamma)$,
where $\calT^\calG_R(\alpha)$ is the translation of $\alpha$ when the
transducer starts from $R$ instead of the initial state
$R_\varepsilon$.

We will not show a direct construction of $\calT^\calG$, when given an
nBPA system $\calG$, but we will show a quickly verifiable ``consistency'' 
condition for any given
transducer $\calT$ that guarantees 
$\calT(\alpha)=\calT(\beta) \Rightarrow \alpha\sim\beta$; moreover,
 $\calT^\calG$ will be shown to satisfy this consistency condition 
 w.r.t. 
$\calG$.

The size of the canonical transducer $\calT^\calG$
is at most exponential in the size of $\calG$
(since $|\calT^\calG_{R}(A)|\leq\norm{A}$, as we will show easily),
and 
we thus have 
a conceptually simple \emph{nondeterministic exponential-time
algorithm deciding 
the nBPA-bbis problem}:
\begin{quote}
	Given a normed BPA system $\calG=(\var,\act,\rules)$
	and $\alpha,\beta\in\var^*$,
guess a transducer $\calT$ of at most
	exponential size (w.r.t. $\calG$), check that $\calT$ 
	is
consistent with $\calG$,
and verify that 
	$\calT(\alpha)=\calT(\beta)$.
\end{quote}\medskip

\noindent In Section~\ref{sec:complexity} we add further remarks on the
construction of $\calT^\calG$ and on the complexity of the nBPA-bbis
problem.

\section{Branching bisimilarity on nBPA via finite
transducers}\label{sec:proof}

\subsection{Normal-form-computing transducers}

By a \emph{transducer} we mean a tuple
$\calT=(Q,\var,\Delta,q_0)$ where $Q$ is a finite set of (control)
states, $\var$ is a finite (input and output) alphabet,
$\Delta$ is a (transition and translation) function
of the type $Q\times \var\longrightarrow Q\times \var^*$,
and $q_0\in Q$ is the initial state.

We view transducers as reading (and writing) 
\emph{from right to left}; we 
 write $q'\lltrans{A}{\gamma}q$ instead of 
$\Delta(q,A)=(q',\gamma)$ to visualize this fact.
The function $\Delta$
is naturally extended to the type
$Q\times \var^*\longrightarrow Q\times \var^*$ by the following 
inductive definition, which uses the ``visual'' notation:
\begin{itemize}
	\item
$q\lltrans{\varepsilon}{\varepsilon}q$ (for each $q\in Q$),
\item
if 
$q'\lltrans{A}{\gamma}q$ and $q''\lltrans{\alpha}{\beta}q'$, then
$q''\lltrans{\alpha A}{\beta\gamma}q$.
\end{itemize}
By 
$\calT_q(\alpha)$ we denote the translation of $\alpha\in\var^*$ 
when starting in $q\in Q$, i.e., the string $\beta$ 
such that $q'\lltrans{\alpha}{\beta}q$ (for some
$q'$); we also use the notation $\calT(\alpha)$ for 
$\calT_{q_0}(\alpha)$.
For each $q\in Q$ we define the equivalence relation
$\equiv^{\calT}_q$ on $\var^*$ as follows:
\begin{center}
$\alpha\equiv^{\calT}_q\beta$ $\iffdef$  $\calT_q(\alpha)=\calT_q(\beta)$; we
put $\equiv^{\calT}=\,\equiv^{\calT}_{q_0}$.
\end{center}
We say that $A\in\var$ is a \emph{$q$-prime} if $\calT_q(A)=A$,
hence if $q'\lltrans{A}{A}q$ for some $q'$.
A string $\alpha\in\var^*$ is a \emph{$q$-normal form} if 
$\alpha=\varepsilon$ or 
$\alpha=A_kA_{k-1}\cdots A_1$ for $k\geq 1$ where
\begin{center}
$q_k\lltrans{A_k}{A_k}q_{k-1}\lltrans{A_{k-1}}{A_{k-1}}q_{k-2}\cdots 
\lltrans{A_3}{A_3}q_2\lltrans{A_2}{A_2}q_1\lltrans{A_1}{A_1}q$ for some $q_1,q_2,\dots,q_k$.
\end{center}
By $\nfT_q$ we denote the set of $q$-normal forms;
hence 
$\varepsilon\in\nfT_q$, and $\beta A\in\nfT_q$ iff $A$ is a $q$-prime
and $\beta$ is a $q'$-normal form for $q'$ satisfying $q'\lltrans{A}{A}q$.
We note that $\alpha\in\nfT_q$ entails
$\calT_q(\alpha)=\alpha$.

A transducer $\calT=(Q,\var,\Delta,q_0)$ 
is a \emph{normal-form-computing
transducer}, an \emph{nfc-transducer} for short, if 
$\calT_q(A)\in\nfT_q$ for all $q\in Q$, $A\in\var$, and
the ``target states'' are the same for both $A$ and $\calT_q(A)$, i.e. 
\begin{equation}\label{eq:nfcondition}
	\textnormal{$q'\lltrans{A}{\gamma}q$ implies
		$q'\lltrans{\gamma}{\gamma}q$ 
		(where $\gamma=\calT_q(A)$).
}		
\end{equation}
For nfc-transducers we thus have  
$\calT_q(\calT_q(\alpha))=\calT_q(\alpha)$ (idempotency), which
also entails that $\alpha\equiv^{\calT}_q\calT_q(\alpha)$;
moreover, the condition~(\ref{eq:nfcondition}) also entails that
$\calT_q(\alpha\beta)=\calT_q(\alpha\,\calT_q(\beta))$.

We note that checking if a given transducer $\calT$ is an
nfc-transducer is straightforward.

\subsection{Nfc-transducers consistent with a BPA system}

In Section~\ref{subsec:canontrans} we will define a canonical
nfc-transducer $\calT^\calG$ for a \emph{normed} BPA system $\calG$; it will turn
out that the branching bisimilarity $\sim$ in $\calL_\calG$ coincides
with the equivalence $\equiv^{\calT^\calG}$.

Here we assume a fixed (general) BPA system $\calG=(\var,\act,\rules)$ 
and a fixed nfc-transducer
$\calT=(Q,\var,\Delta,q_0)$; we aim to find a suitable 
 condition guaranteeing
that  the equivalence $\equiv^{\calT}$ (on the set
$\var^*$) is a branching
bisimulation in the LTS
$\calL_\calG=(\var^*,\act,(\trans{a})_{a\in\act})$.

A natural idea is to require that for every action $a\in\act$
(including the case $a=\tau$) the processes $\alpha$ and $\calT(\alpha)$ 
yield 
the same normal forms of the results of ``long moves''
$\trans{\tau}\trans{\tau}\cdots\trans{\tau}\trans{a}$
where the (maybe empty) $\tau$-prefix is bound to go inside the
equivalence class
$[\alpha]_{\equiv^{\calT}}$ (which is the same as
$[\calT(\alpha)]_{\equiv^{\calT}}$), and the final $\trans{a}$-step might be missing
when $a=\tau$. 
We formalize this idea by Def.~\ref{def:constrans}, after 
we introduce  the ``long
moves''  $\strans{a}_q$, relativized w.r.t. the states $q\in Q$.

\medskip

For our fixed  $\calG$ and 
$\calT$ we write $\alpha\strans{a}_q\beta$, where 
$\alpha,\beta\in\var^*$, $q\in Q$, and $a\in\act$, if 
\begin{itemize}
	\item		
either	$a=\tau$ and $\beta=\calT_q(\alpha)$, 
	\item
or		there are $\alpha_1,\alpha_2,\dots,\alpha_k$ (for some $k\geq
		0$) and $\beta'$ such that

\[\alpha=\alpha_0\trans{\tau}\alpha_1\trans{\tau}\cdots\trans{\tau}\alpha_k\trans{a}\beta'\qquad\mbox{in
    $\calL_\calG$,}
\]
\[\calT_q(\alpha_0)=\calT_q(\alpha_1)=\cdots=\calT_q(\alpha_k)
\qquad\mbox{and}\qquad\calT_q(\beta')=\beta\ .
\]
\end{itemize}
Hence $\alpha\strans{a}_q\beta$ entails that $\beta \in\nfT_q$
($\beta$ is a $q$-normal form).
In particular we have $\varepsilon\strans{\tau}_q\varepsilon$.
We define the equivalences $\approx_q$
as follows:
\begin{center}
$\alpha_1\approx_q\alpha_2$\ \ $\iffdef$\ \
$\forall a\in\act: \{\beta\mid \alpha_1\strans{a}_q\beta\}=
\{\beta\mid \alpha_2\strans{a}_q\beta\}$.
\end{center}

Now the announced definition follows; it also uses the fact
that $q'\lltrans{A}{\varepsilon} q$ implies $q'=q$ for
nfc-transducers (by the condition~(\ref{eq:nfcondition})).

\begin{defi}\label{def:constrans}
An  \emph{nfc-transducer}
$\calT=(Q,\var,\Delta,q_0)$ is
\emph{consistent 
with a BPA system} $\calG=(\var,\act,\rules)$ 
if the following three conditions hold.
\begin{enumerate}
	\item
		$A\approx_{q_0}\varepsilon$
		if
		$\calT_{q_0}(A)=\varepsilon$ (i.e., if
		$q_0\lltrans{A}{\varepsilon} q_0$);
	\item		
$A\approx_q\calT_q(A)$ if
	$\calT_q(A)\neq\varepsilon$ 
(hence $q'\lltrans{A}{\beta}q$ where $\beta\neq\varepsilon$ entails
$A\approx_q\beta$);
\item	
	$AC\approx_q C$
	if $\calT_q(AC)=\calT_q(C)=C$
	(i.e., if
	$q'\lltrans{A}{\varepsilon} q'\lltrans{C}{C}q$ 
	for some $q'$).
\end{enumerate}
\end{defi}

\begin{lem}\label{lem:constrans}
\ \hfill
\begin{enumerate}
\item
There is a polynomial algorithm checking if a given nfc-transducer 
$\calT$ is consistent with a given BPA system $\calG$.
\item
If an nfc-transducer $\calT$ is consistent with a BPA system
$\calG$,
then $\equiv^\calT$ 
is a branching bisimulation in $\calL_{\calG}$.
\end{enumerate}
\end{lem}	
\proof\ \\
	1.
We assume an nfc-transducer $\calT=(Q,\var,\Delta,q_0)$ and a BPA
system $\calG=(\var,\act,\calR)$.
For any $q\in Q$, we put
$\eimage_q=\{X\in\var\mid\calT_q(X)=\varepsilon\}$,
and we define the set $\eras_q\subseteq\var$
(of \emph{silently $q$-erasable variables})
inductively:
\begin{center}
$X\in \eras_q$ if $X\in\eimage_q$
and 
there is a rule $X\trans{\tau}\gamma$ in $\calR$ where $\gamma\in (\eras_q)^*$.
\end{center}
Using dynamic programming, the sets  $\eras_q$ are quickly constructible
for all $q\in Q$. (In the first step we find $X\in\eimage_q$ for
which  $X\trans{\tau}\varepsilon$ is a rule in $\calR$; if there are
no such $X$, then $\eras_q=\emptyset$.)

\medskip

It is easy to verify that the following ``axioms and
	deduction rules'' i) -- v)
 characterize when we have $\alpha\strans{a}_q\beta$. 
(We omit $\calT(\alpha)$ in the notation
$q'\lltrans{\alpha}{\calT(\alpha)}q$ when
unimportant.)
\begin{enumerate}[label={\roman*)}]
	\item $\alpha\strans{\tau}_q\calT_q(\alpha)$ (for all
		$\alpha\in\var^*$ and $q\in Q$);
	\item
		if $A\trans{a}\delta$ is a rule in $\calR$,
		then $A\strans{a}_{q}\calT_{q}(\delta)$;
\item
	if $A\trans{\tau}\delta$ is a rule in $\calR$,
	$\calT_q(A)=\calT_q(\delta)$, and $\delta\strans{a}_q\beta$,
	then $A\strans{a}_q\beta$;
	\item		
if $q''\lltrans{A}{}q'\lltrans{\gamma}{\delta}q$ and
$A\strans{a}_{q'}\beta$, then 
$A\gamma\strans{a}_{q}\beta\delta$;
\item
if $q'\lltrans{A}{\varepsilon}q'\lltrans{\gamma}{}q$, 
$A\in\eras_{q'}$, and $\gamma\strans{a}_{q}\beta$, then
$A\gamma\strans{a}_{q}\beta$.
\end{enumerate}
We say that a \emph{string} $\alpha\in\var^*$ is \emph{basic} if it is
just one
variable ($\alpha\in\var$) or it is a suffix of the right-hand side $\delta$ in a rule
	$A\trans{a}\delta$ in $\calR$; hence the number of basic
	strings is no bigger than a standard size of $\calG$.
	We say that $\alpha\strans{a}_q\beta$ is a \emph{basic move} if
$\alpha$ is a basic string. 
Any basic move  $\alpha\strans{a}_q\beta$ can be derived in the 
``deduction system'' i) -- v) either by using an axiom i) or ii), or
by using another basic move with a shorter derivation (in the rules
iii) -- v)).
Hence if we apply  i) -- v)
only to basic strings 
(i.e., we use i) only if the respective $\alpha$ is basic,
and we use iv) or v) only if $A\gamma$ is basic) 
iteratively as long as new basic moves are being derived,
we get all basic moves.
Moreover, if  $\alpha\strans{a}_q\beta$ is a basic move, then 
$\beta$ is either $\calT_q(\alpha)$ or
of the form $\calT_{q'}(\delta)\,\gamma$ for some $q'\in Q$ 
where $\gamma$ is a
suffix of $\calT_q(\alpha)$ and $\delta$ is the right-hand side of a rule
in $\calR$ (and $q'\lltrans{\gamma}{\gamma}q$); this claim also
follows
inductively, when inspecting the rules i)--v).
There are thus only polynomially many basic moves (in the size of
$\calG$ and $\calT$).

The above observations immediately yield a polynomial algorithm 
(in the size of $\calG$ and
$\calT$) that 
constructs all basic moves.
A polynomial check of consistency of $\calT$ with $\calG$ will be thus clear 
after we show that also
non-basic moves of the type $\calT_q(A)\strans{a}_q\gamma$ 
and 
$AC\strans{a}_q\gamma$ where $\calT(AC)=\calT(C)=C$ can be easily 
 constructed, when basic moves are
given.
Let $\calT_q(A)=B\beta$, where 
$q''\lltrans{B}{B}q'\lltrans{\beta}{\beta}q$ (recall that
$\calT_q(A)$
is a $q$-normal form);
then $B\beta\strans{a}_q\gamma$ iff $\gamma=\gamma'\beta$ and
$B\strans{a}_{q'}\gamma'$.
If $q'\lltrans{A}{\varepsilon}q'\lltrans{C}{C}q$
then $AC\strans{a}_{q}\gamma$ iff $\gamma=\gamma'C$ and 
$A\strans{a}_{q'}\gamma'$ or $A\in\eras_q$ and
$C\strans{a}_{q}\gamma$.

\bigskip

2.
	Let 
 $\calT=(Q,\var,\Delta,q_0)$ be an nfc-transducer that is consistent
 with a given BPA system $\calG=(\var,\act,\rules)$.
We will first show that $\equiv^{\calT}$ is a branching bisimulation
in $\calL_\calG$ \emph{when assuming
$\alpha\approx\calT(\alpha)$} for all
$\alpha\in\var^*$, where $\approx$ stands for
$\approx_{q_0}$;
this assumption
will be proven afterwards.

Let us consider some
 $\alpha\equiv^{\calT}\beta$ and a transition $\alpha\trans{a}\alpha'$;
 we thus have  $\alpha\strans{a}\calT(\alpha')$ where $\strans{a}$
 stands for  $\strans{a}_{q_0}$. 
  If $a=\tau$ and $\calT(\alpha)=\calT(\alpha')$, then 
$\alpha'\equiv^\calT\beta$; so we further suppose that $a\neq\tau$
or $\calT(\alpha)\neq\calT(\alpha')$.
 Since
 $\alpha\approx\calT(\alpha)=\calT(\beta)\approx\beta$, 
 we must also have  $\beta\strans{a}\calT(\alpha')$.
Hence we have 
$\beta=\beta_0\trans{\tau}\beta_1\trans{\tau}\cdots\trans{\tau}\beta_k\trans{a}\beta'$
for some $k\geq 0$ where 
$\calT(\beta_0)=\calT(\beta_1)=\cdots=\calT(\beta_k)$
and $\calT(\beta')=\calT(\alpha')$. Therefore
$\alpha\equiv^\calT\beta_i$ for all $i\in[0,k]$, and
$\alpha'\equiv^\calT\beta'$. 
We have thus verified that  
$\equiv^{\calT}$ is indeed a branching bisimulation.

It remains to prove that $\alpha\approx\calT(\alpha)$.
We proceed by induction on $|\alpha|$. If $\alpha=\varepsilon$, then
the claim is trivial. We now assume $\alpha=A\beta$ where
$q'\lltrans{A}{\calT_q(A)}q\lltrans{\beta}{\calT(\beta)}q_0$ and
$\beta\approx\calT(\beta)$. The fact
$q\lltrans{\calT(\beta)}{\calT(\beta)}q_0$
(due to the properties of \emph{nfc}-transducers) then implies 
$A\beta\approx A\,\calT(\beta)$ (as can be verified by iv) and v) in
the above ``deduction system''); for establishing 
$A\beta\approx\calT(A\beta)$
it thus suffices to show that
$A\,\calT(\beta)\approx \calT_q(A)\,\calT(\beta)$ (recall that 
$\calT(A\beta)=\calT_q(A)\,\calT(\beta)$).
If $\calT_q(A)\neq\varepsilon$, then 
this follows from $A\approx_q\calT_q(A)$ (cf.
Def.~\ref{def:constrans}(2)).
Hence we further assume $\calT_q(A)=\varepsilon$.

If $\calT(\beta)=\varepsilon$ (hence
$q_0\lltrans{A}{\varepsilon}q_0\lltrans{\beta}{\varepsilon}q_0$),
then we need to show 
that $A\approx\varepsilon$ (i.e. $A\approx_{q_0}\varepsilon$);
this holds by Def.~\ref{def:constrans}(1).
If $\calT(\beta)=C\delta$, hence 
$q\lltrans{A}{\varepsilon}q\lltrans{C}{C}q''\lltrans{\delta}{\delta}q_0$,
then $AC\delta\approx C\delta$ follows from $AC\approx_{q''}C$, which
 holds by Def.~\ref{def:constrans}(3).
\qed

\subsection{Canonical transducers}\label{subsec:canontrans}

Given a \emph{normed} BPA system 
$\calG=(\var,\act,\rules)$, we now show a (non-effective) construction
of a canonical transducer $\calT^\calG$.
It will turn out
that $\calT^\calG$ is an nfc-transducer that is consistent with
$\calG$ (hence $\equiv^{\calT^\calG}\subseteq\sim$) and for which 
 $\sim\subseteq\equiv^{\calT^\calG}$;
 hence $\alpha\sim\beta$ in $\calL_\calG$
 iff $\calT^\calG(\alpha)=\calT^\calG(\beta)$.
We will also get an exponential bound on the size of
$\calT^\calG$ (in the size of $\calG$).
These facts will immediately entail a $\NEXPTIME$ upper bound for the
branching bisimilarity problem for normed BPA systems. (We have
already touched on this in Section~\ref{sec:overview}, and
some further
remarks on the complexity are in Section~\ref{sec:complexity}.)

In the definition of the transducer $\calT^\calG$ we also use 
the following technical notions.
For $\gamma\in\var^*$ we put 
$$R_\gamma=\{X\in\var\mid X\gamma\sim\gamma\}.$$
Each $X\in R_\gamma$ is
called a \emph{redundant variable w.r.t.} $\gamma$. 
We say that the prefix $\alpha$ of 
$\alpha\gamma\in\var^*$ is \emph{redundancy-free} if it cannot be written 
as $\alpha=\delta X\beta$ where $X\beta\gamma\sim\beta\gamma$.

To make $\calT^\calG$ unique (though this is not crucial), we
also assume a linear order on the set $\var$; we say that
$\alpha\in\var^*$ is \emph{lexicographically smaller} than
$\beta\in\var^*$ if $\alpha$ is a proper suffix of $\beta$, or  
if $\alpha=\alpha'A\gamma$, $\beta=\beta'B\gamma$ and $A$ is less than
 $B$ in the order on $\var$. (We reflect our right-to-left transducers in this
definition.)

We first state the following definition and then we discuss 
its soundness, which is based on the assumption that 
$\calG$ is normed.

\begin{defi}\label{def:canontrans}
For a normed BPA system 
$\calG=(\var,\act,\rules)$, where $\var$ is linearly ordered,
we define the \emph{canonical transducer}  
$\calT^\calG=(Q,\var,\Delta,q_0)$ by the following three points.
	\begin{enumerate}[label={\roman*)}]
	\item
$Q=\{R_\gamma\mid \gamma\in\var^*\}$.
(Hence each state is the set of redundant variables
 w.r.t. some $\gamma$.)
\item 
The initial state $q_0$ is the set $R_\varepsilon$
(i.e. the set $\{X\in\var\mid X\sim\varepsilon\}$, which might be empty).
	\item		
For each $R_\gamma\in Q$ and each $A\in\var$ we put
		$\Delta(R_\gamma,A)=(R_{A\gamma},\alpha)$, 
 which is denoted	as 
 \\
$R_{A\gamma}\lltrans{A}{\alpha}R_\gamma$, where $\alpha$ satisfies
		the conditions 
		\begin{enumerate}[label={\alph*)}]
			\item
$\alpha\gamma\sim A\gamma$, 
\item
$\alpha$ is 
a redundancy-free prefix of $\alpha\gamma$,
		\end{enumerate}				
and is lexicographically smallest among the longest strings
		satisfying a) and b).

\end{enumerate}		
\end{defi}

\noindent The soundness of the definition can be shown by the facts
established already in~\cite{DBLP:conf/icalp/Fu13}; a crucial fact 
is that
$R_\gamma=R_\delta$ implies 
$\alpha\gamma\sim\beta\gamma$ $\ \Leftrightarrow\ $
$\alpha\delta\sim\beta\delta$. 
To be self-contained, we also prove these facts
(by Prop.~\ref{prop:Requiv}), and then we show
the soundness (as a part of Theorem~\ref{th:canontrans}).

We fix a normed BPA system
$\calG=(\var,\act,\calR)$, and we first define  
the ``relative'' equivalences $\alpha\sim_R\beta$ 
and the ``relative'' cc-norms $\bnorm{\alpha}_R$
for all
$R\subseteq \var$,
via the LTSs $\calL_{\calG,R}$; for a fixed set $R\subseteq\var$ we
stipulate: 
\begin{itemize}
	\item	
		The LTS $\calL_{\calG,R}$ arises from
		$\calL_\calG=(\var^*,\act,(\trans{a})_{a\in\act})$ by declaring all
$\alpha\in R^*$ to be silent states; technically we simply 
		remove
		all their outgoing transitions.
Hence $\alpha\in R^*$
		satisfies $\alpha\sim\varepsilon$ in
		$\calL_{\calG,R}$.
	\item
		$\alpha\sim_R\beta$ $\ \iffdef\ $
$\alpha\sim\beta$ in 	$\calL_{\calG,R}$.
\item
	$\bnorm{\alpha}_R$ is equal to $\bnorm{\alpha}$
		in 	$\calL_{\calG,R}$.
\end{itemize}

\begin{rem}
Unlike in~\cite{DBLP:conf/icalp/Fu13},
the definition is not restricted to $R=R_\gamma$ for $\gamma\in\var^*$,
and the claims that we derive for $R_\gamma$ can be naturally extended to the
general cases $R\subseteq \var$. Similarly we could define 
the states of $\calT^\calG$
to be all sets $R\subseteq \var$ (not only those reachable from
$R_\varepsilon$). 
Additional remarks are given in 
Section~\ref{sec:complexity}.
\end{rem}

Now we note a few facts that already appeared
in~\cite{DBLP:conf/icalp/Fu13}. \newpage

\begin{prop}\label{prop:Requiv}
For any nBPA system $\calG$ the following claims hold:
	\begin{enumerate}
		\item	$\gamma\sim\delta$ implies
			$\alpha\gamma\sim\alpha\delta$;		
		\item
	$\alpha\gamma\sim\gamma$ iff $\alpha\in (R_\gamma)^*$;
		\item
			$\alpha\gamma\sim\beta\gamma$ iff
	$\alpha\sim_{R_\gamma}\beta$. 
	\end{enumerate}
	\end{prop}
\proof	
We assume an nBPA system $\calG=(\var,\act,\calR)$.
\begin{enumerate}
	\item If $\gamma\sim\delta$, then the set
	$\calB=\,\,\sim \cup\,\, \{(\alpha\gamma,\alpha\delta) \mid
	\alpha\in\var^*\}$ can be easily verified to be
	a branching bisimulation (and thus $\calB=\,\sim$ in
	$\calL_\calG$).
	
\medskip

\item
If $\alpha=\alpha'Y$ where $Y\in R_\gamma=\{X\mid X\gamma\sim \gamma\}$,
	then $\alpha\gamma=\alpha'Y\gamma\sim \alpha'\gamma$ (by 
	1.); using this fact repeatedly,
	$\alpha\in (R_\gamma)^*$ entails $\alpha\gamma\sim\gamma$.

	Suppose $\alpha\not\in(R_\gamma)^*$, hence 
	$\alpha=\alpha'Y\alpha''$ where $\alpha''\in (R_\gamma)^*$ and
	$Y\not\in R_\gamma$; thus $\alpha\gamma\sim\alpha'Y\gamma$.
Since $Y\gamma\not\sim\gamma$, we have 
$\bnorm{Y\gamma}>\bnorm{\gamma}$ (since any path $Y\gamma\trans{u}\gamma$
contains at least one class-changing transition); this entails
	$\bnorm{\alpha'Y\gamma}\geq\bnorm{Y\gamma}>\bnorm{\gamma}$,
	and thus
	$\alpha'Y\gamma\not\sim\gamma$ (by Observation~\ref{obs:ccnorm}).
Since $\alpha\gamma\sim\alpha'Y\gamma$, we get
	$\alpha\gamma\not\sim\gamma$.

	\medskip

	\item
\begin{enumerate}[label=\alph*)]% a)
\item We first show
the implication 
$\alpha\gamma\sim\beta\gamma\ \Rightarrow\
\alpha\sim_{R_\gamma}\beta$.
	 This 
will be clear when we show that for any 	
	$\gamma\in\var^*$ the set
\begin{center}
	$\calB=\{(\alpha,\beta)\mid
	\alpha\gamma\sim\beta\gamma\}$
\end{center}	
is a branching bisimulation in  
$\calL_{\calG,R_{\gamma}}$. 
Let $(\alpha,\beta)\in \calB$ and $\alpha\trans{a}\alpha'$ in
	$\calL_{\calG,R_\gamma}$; we will show that the move $\alpha\trans{a}\alpha'$
	can be matched from $\beta$ in $\calL_{\calG,R_{\gamma}}$.
We note that $\alpha\not\in (R_\gamma)^*$
	(since it has an outgoing transition 
	in $\calL_{\calG,R_{\gamma}}$) and thus 
$\alpha\gamma\not\sim\gamma$ (by 2.);
this also entails $\beta\gamma\not\sim\gamma$
	since
 $\alpha\gamma\sim\beta\gamma$. We also have
the move $\alpha\gamma\trans{a}\alpha'\gamma$ in
$\calL_{\calG}$.
\begin{itemize}
	\item
If $a=\tau$ and $\alpha'\gamma\sim\alpha\gamma$, hence also 
$\alpha'\gamma\sim\beta\gamma$,	then $(\alpha',\beta)\in \calB$.	
\item
If $a\neq\tau$ or $\alpha'\gamma\not\sim\alpha\gamma$, then in 
$\calL_{\calG}$
we must
have 
$\beta\gamma=\delta_0\trans{\tau}\delta_1\cdots
\trans{\tau}\delta_k\trans{a}\delta$
where $\alpha\gamma\sim \beta\gamma\sim \delta_i$ for all $i\in[0,k]$ and
$\alpha'\gamma\sim\delta$. Since $\beta\gamma\not\sim\gamma$, we have
 $\delta_i\not\sim\gamma$ for all $i\in[0,k]$; this entails that for
		each $i\in[0,k]$ we have
		$\delta_i=\beta_i\gamma$ where $\beta_i\not\in
		(R_\gamma)^*$,  
$\delta=\beta'\gamma$, and  
$\beta=\beta_0\trans{\tau}\beta_1\cdots
\trans{\tau}\beta_k\trans{a}\beta'$ is a path 
		in $\calL_{\calG,R_\gamma}$. Hence 
		$\alpha\gamma\sim\beta_i\gamma$
for all $i\in[0,k]$, and $\alpha'\gamma\sim\beta'\gamma$;
therefore $(\alpha,\beta_i)\in\calB$
for all $i\in[0,k]$, and $(\alpha',\beta')\in\calB$.
\end{itemize}			

\item %b)
Now we show the implication 
	$\alpha\sim_{R_\gamma}\beta\ \Rightarrow\
	\alpha\gamma\sim\beta\gamma$. This 
will be clear when we show that for any 	
	$\gamma\in\var^*$ the set
\begin{center}
	$\calB=\,\,\sim\cup\,\,\{(\alpha\gamma,\beta\gamma)\mid
	\alpha\sim_{R_\gamma}\beta\}$
\end{center}	
is a branching bisimulation in  
$\calL_{\calG}$ (which implies $\calB=\,\sim$ in $\calL_\calG$). 
	It suffices to show that for any $(\delta_1,\delta_2)\in\calB$
	any 
	move $\delta_1\trans{a}\delta$ can be matched  
 from $\delta_2$. If $\delta_1\sim \delta_2$, then
	this follows from the definition of $\sim$. Hence it suffices
	to consider the case 
 $\delta_1=\alpha\gamma$ and $\delta_2=\beta\gamma$
where $\alpha\sim_{R_\gamma}\beta$.
	\begin{itemize}	
		\item
			If $\alpha\in (R_\gamma)^*$, then
			$\alpha\gamma\sim\gamma$ and
			$\alpha\sim_{R_\gamma}\varepsilon\sim_{R_\gamma}\beta$; both
			$\alpha$ and 
			$\beta$ are silent states in
			$\calL_{\calG,R_\gamma}$. This entails that
			either $\beta\in (R_\gamma)^*$, or any move 
			$\beta\trans{b}\beta'$ in 
			$\calL_{\calG,R_\gamma}$ satisfies that $b=\tau$
			and that $\beta'$ is silent in
			$\calL_{\calG,R_\gamma}$;
			since $\calG$ is normed, we must also have
			$\beta\trans{u}\beta'$ where $u\in\{\tau\}^*$
			and $\beta'\in(R_\gamma)^*$.
			It follows 
			that $\beta\gamma\sim\gamma$ (since
			the set
			$\sim\cup\,\{(\beta'\gamma,\gamma)\mid\beta'$
			is silent in  $\calL_{\calG,R_\gamma}\}$ is a
			branching bisimulation); hence 
			$\beta\in(R_\gamma)^*$, in fact (by 2.). We thus have 
			the case $\delta_1\sim\delta_2$ 
			(since
			$\alpha\gamma\sim\gamma\sim\beta\gamma$);
			the move $\delta_1\trans{a}\delta$ can be
			thus matched from $\delta_2$.
\item
	If $\alpha\not\in (R_\gamma)^*$, then the move
	$\delta_1\trans{a}\delta$, i.e.
			$\alpha\gamma\trans{a}\delta$,
			can be presented
			as 
			$\alpha\gamma\trans{a}\alpha'\gamma$ where 
$\alpha\trans{a}\alpha'$ in $\calL_{\calG,R_\gamma}$.
If $a=\tau$ and $\alpha'\sim_{R_\gamma}\alpha$, then we have
$(\delta,\delta_2)=
(\alpha'\gamma,\beta\gamma)\in\calB$.
Now assume 	 $a\neq\tau$ or $\alpha'\not\sim_{R_\gamma}\alpha$.
Since $\alpha\sim_{R_\gamma}\beta$, in $\calL_{\calG,R_\gamma}$ we must have 
$\beta=\beta_0\trans{\tau}\beta_1\cdots\trans{\tau}\beta_k\trans{a}\beta'$
where $\alpha\sim_{R_\gamma}\beta_i$ for all $i\in[0,k]$ and
$\alpha'\sim_{R_\gamma}\beta'$. But then in $\calL_{\calG}$ we have
$\delta_2=\beta_0\gamma\trans{\tau}\beta_1\gamma\cdots\trans{\tau}\beta_k\gamma\trans{a}\beta'\gamma$
			where $(\alpha\gamma,\beta_i\gamma)\in\calB$ for all 
			$i\in[0,k]$ and
			$(\alpha'\gamma,\beta'\gamma)\in\calB$. Hence
			the move $\delta_1\trans{a}\delta$
			(i.e. $\alpha\gamma\trans{a}\alpha'\gamma$)
can be matched from $\delta_2=\beta\gamma$.\qedhere
	\end{itemize}
\end{enumerate}
\end{enumerate}

\noindent We now prove the announced properties of $\calT^\calG$
(from Def.~\ref{def:canontrans}).

\begin{thm}\label{th:canontrans}
 For any normed BPA system $\calG$, the canonical transducer $\calT^\calG$
 has the following properties:
	\begin{enumerate}
		\item
$\calT^\calG$ is an nfc-transducer 
that is consistent with $\calG$.
		\item
			$\equiv^{\calT^\calG}=\,\sim$ (i.e., 
			$\calT^\calG(\alpha)=\calT^\calG(\beta)$ iff
			$\alpha\sim\beta$ in $\calL_\calG$).
		\item
The size of $\calT^{\calG}$ is bounded by an exponential function
of the size of $\calG$.
\end{enumerate}
\end{thm}
\proof
Let $\calG=(\var,\act,\calR)$ be an nBPA system and
let $\calT^\calG=(Q,\var,\Delta,q_0)$ be 
as in Def.~\ref{def:canontrans}.

\medskip
\begin{enumerate}
\item%	1.
 (First part.) We now show that $\calT^\calG$ is an
	nfc-transducer; the consistency with $\calG$ is shown in the second part, after
	the point (2) is established.

We first need to show that the function $\Delta$, presented by
four-tuples $R_{A\gamma}\lltrans{A}{\alpha}R_{\gamma}$, is defined
	soundly. Let us assume $R_\gamma=R_\delta$, hence
	$\sim_{R_\gamma}=\,\sim_{R_\delta}$.
		By Prop.~\ref{prop:Requiv}(3) we deduce that $XA\gamma\sim
		A\gamma$ iff $XA\delta\sim
		A\delta$; hence $R_{A\gamma}=R_{A\delta}$.
Similarly we deduce that $\alpha\gamma\sim A\gamma$ iff 
	$\alpha\delta\sim A\delta$, and that 
$\alpha$ is a redundancy-free prefix of $\alpha\gamma$ iff 
$\alpha$ is a redundancy-free prefix of $\alpha\delta$.
	Hence the strings $\alpha$ satisfying a) and b) 
	in Def.~\ref{def:canontrans}(iii) are determined by
	the set $R_\gamma$. The set of such strings is nonempty 
	(since it contains $\alpha=A$ or $\alpha=\varepsilon$); once
	we show that this set is finite, the soundness of $\Delta$ is
	clear. 
	The finitiness 
	follows from the fact that $\alpha\gamma\sim A\gamma$ entails 
	$\bnorm{\alpha\gamma}=\bnorm{A\gamma}$ 
	(Observation~\ref{obs:ccnorm}), and from the obvious fact 
	that $\bnorm{\alpha\gamma}\geq |\alpha|+\bnorm{\gamma}$ 
	when $\alpha$ is a
	redundancy-free prefix of $\alpha\gamma$. (We have already
	observed that
	$X\beta\not\sim\beta$ entails that any path $X\beta\trans{u}\beta$
	has at least one class-changing transition.)

Hence $\calT^\calG$ is indeed a transducer. 
	We show that it is an
	nfc-transducer, i.e.,
$R'\lltrans{A}{\alpha}R$ implies that $\alpha$ is an $R$-normal form
	and
	$R'\lltrans{\alpha}{\alpha}R$.

	Let us consider $R_{A\gamma}\lltrans{A}{\alpha}R_\gamma$.
	By definition of $\calT^\calG$, and by
	Prop.~\ref{prop:Requiv}(3), the string $\alpha$ is
	lexicographically smallest among the longest strings that
	satisfy $\alpha\sim_{R_\gamma}A$ and are $R_\gamma$-redundancy
	free, by which we mean that $\alpha=\delta X\beta$ entails
	$X\beta\not\sim_{R_\gamma}\beta$.
	This obviously entails that $\alpha=\varepsilon$ iff $A\in
	R_{\gamma}$ (by recalling Prop.~\ref{prop:Requiv}(2)); in this
	case we have $R_{\gamma}\lltrans{A}{\varepsilon}R_\gamma$.
	
	We thus further suppose $A\not\in R_\gamma$.
Since $A\gamma\sim\alpha\gamma$, we have
$XA\gamma\sim X\alpha\gamma$ (recall Prop.~\ref{prop:Requiv}(1)); therefore
	$XA\gamma\sim A\gamma$ iff $X\alpha\gamma\sim \alpha\gamma$.
Hence
	$R_{\alpha\gamma}=R_{A\gamma}$, and we thus have 
$R_{A\gamma}\lltrans{\alpha}{}R_\gamma$.
It remains to show that $\alpha$ is 
an $R_\gamma$-normal form.
For the sake of contradiction we suppose that it is not
the case;
hence we have
$\alpha=\alpha'B A_\ell
	A_{\ell-1}\cdots A_2A_1$ for some $\ell\geq 0$ where
	\begin{center}
	$R_{\alpha\gamma}\lltrans{\alpha'}{\beta'}R_{\ell+1}\lltrans{B}{\beta}R_\ell\lltrans{A_\ell}{A_\ell}\cdots
	R_2\lltrans{A_2}{A_2}R_{1}\lltrans{A_1}{A_1}R_\gamma$ and
	$\beta\neq B$. 
	\end{center}
By the definition of $\calT^\calG$ we have $B\sim_{R_\ell}\beta$ and
$\beta\neq\varepsilon$ (we have $B\not\in R_\ell$ since $\alpha$ is
$R_\gamma$-redundancy free), which entails that 
$|\alpha'\beta A_\ell\cdots A_2A_1|>|\alpha|$ or
$|\alpha'\beta A_\ell\cdots A_2A_1|=|\alpha|$ and 
$\alpha'\beta A_\ell\cdots A_2A_1$ is 
lexicographically
smaller than $\alpha$.
The fact $\beta\sim_{R_\ell}B$ entails
$\beta A_\ell\cdots A_1\gamma\sim BA_\ell\cdots A_1\gamma$, and thus
$\alpha'\beta A_\ell\cdots A_1\gamma\sim\alpha'BA_\ell\cdots
A_1\gamma=\alpha\gamma$ (using Prop.~\ref{prop:Requiv}(1,3)).
This forces us to conclude that $\alpha'\beta A_\ell\cdots A_1$ is not
$R_\gamma$-redundancy free (due to the choice of $\alpha$ in
$\calT^\calG$).
But this is impossible, since
$\beta$ is $R_\ell$-redundancy free, $\alpha'$ is
$R_{\ell+1}$-redundancy free,
and $R_{\ell+1}\lltrans{\beta}{}R_\ell$ (since $\beta\sim_{R_\ell}B$
entails $X\beta\sim_{R_\ell}\beta$ iff $XB\sim_{R_\ell}B$).

%\bigskip	

\item We will show the following (more general) claim for the
nfc-transducer $\calT^\calG$:
\begin{equation}\label{eq:simisequiv}
\textnormal{
$\alpha\sim_R\beta$ iff
$\calT^\calG_R(\alpha)=\calT^\calG_R(\beta)$ (for any $R\in Q$).
}
\end{equation}	
We first note the fact that
 \begin{equation}\label{eq:simwithnf}
	 \textnormal{ $\alpha\sim_R\calT^\calG_R(\alpha)$ (for all $\alpha\in\var^*$ and
	 $R\in Q$),} 
\end{equation}
using an 
induction on $|\alpha|$. For $\alpha=\varepsilon$ the fact
($\varepsilon\sim_R\varepsilon$) is trivial.
For $\alpha =\alpha'A$ we have
	$R''\lltrans{\alpha'}{\calT^\calG_{R'}(\alpha')}R'\lltrans{A}{\calT^\calG_R(A)}R$
	where 
$A\sim_R\calT^\calG_R(A)$ by the definition of $\calT^\calG$
and 
$\alpha'\sim_{R'}{\calT^\calG_{R'}(\alpha')}$ by the induction
hypothesis. Hence $\alpha'A\sim_R\calT^\calG_R(\alpha'A)$, by
applying Prop.~\ref{prop:Requiv}(3). 

\medskip

The ``if-direction'' of~(\ref{eq:simisequiv}) 
thus follows (since
$\alpha\sim_R\calT^\calG_R(\alpha)=\calT^\calG_R(\beta)\sim_R\beta$
implies $\alpha\sim_R\beta$).

We now show the ``only-if-direction'' of~(\ref{eq:simisequiv}).
For the sake of contradiction, suppose there are $\alpha\sim_R\beta$
for which $\calT^\calG_R(\alpha)\neq\calT^\calG_R(\beta)$. 
By~(\ref{eq:simwithnf}) 
we deduce that there are two different $R$-normal forms
$\alpha,\beta$ such that 
$\alpha\sim_R\beta$; let us consider such  $\alpha,\beta$.
We thus have $\alpha=\alpha'A\gamma$, $\beta=\beta' B\gamma$ 
where $A\neq B$; hence 
$\alpha'A\sim_{R'}\beta'B$ where
$R'\lltrans{\gamma}{\gamma}R$, and $\alpha'A$, $\beta'B$ are
$R'$-normal forms.
Hence we immediately choose some $R\in Q$ and two $R$-normal forms 
$\alpha A$, $\beta B$ where $A\neq B$ and $\alpha A\sim_R\beta B$;
w.l.o.g. we assume $\bnorm{A}_R\geq\bnorm{B}_R$.

We now consider a path $\alpha A\trans{u} \gamma A$ in $\calL_{\calG,R}$ where 
$\bnorm{\gamma A}_R=\bnorm{A}_R$ 
and the $R$-cc-norms $\bnorm{.}_R$ of all processes on this path before  $\gamma A$
are bigger than $\bnorm{A}_R$. (We can have $u=\varepsilon$ and 
$\gamma=\alpha$; in this case $\alpha=\varepsilon$ since otherwise
$\bnorm{\alpha A}_R>\bnorm{A}_R$ due to $R$-redundancy freeness of
$\alpha A$, which is an $R$-normal form.)

We recall that $\gamma_1\sim_R\gamma_2$ implies
$\bnorm{\gamma_1}_R=\bnorm{\gamma_2}_R$ (by
Observation~\ref{obs:ccnorm}).
Since $\alpha A\sim_R\beta B$, the path $\alpha A\trans{u} \gamma A$
must have a matching path 
$\beta B\trans{v} \delta'$ where 
$\gamma A\sim_R \delta'$ 
(hence $\bnorm{\delta'}_R=\bnorm{\gamma A}_R=\bnorm{A}_R\geq
\bnorm{B}_R$) and all processes on the path before  $\delta'$
have the $R$-cc-norms bigger than $\bnorm{A}_R$; necessarily $\delta'=\delta
B$ for some $\delta$.

We now derive a contradiction. Since $\bnorm{\gamma A}_R=\bnorm{A}_R$,
we have $\gamma A\sim_R A$ (there must be a path $\gamma A\trans{w} A$
with no class-changing transition), and thus $A\sim_R \delta B$.
We have $\calT^\calG_R(A)=A$ ($A$ is an $R$-prime 
since $\alpha A$ is an $R$-normal form)
and $A\sim_R \delta B\sim_R \calT^\calG_R(\delta B)=\delta'' B$
(for some $\delta''$; $B$ is also an $R$-prime since $\beta B$ is an
$R$-normal form). 
If
$\delta''\neq\varepsilon$, 
then $A$ is not a longest $R$-redundancy-free string from the class
$[A]_{\sim_R}$  (which contradicts with $\calT^\calG_R(A)=A$);
 if $\delta''=\varepsilon$, then $A\sim_R B$ and one of
$\calT^\calG_R(A)=A$, 
$\calT^\calG_R(B)=B$ violates the ``lexicographically smallest''
condition.

%\bigskip 

\item[(1)]%1. 
(Second part.) We show that
$\calT^\calG$ is consistent with $\calG$.
We have to show $A\approx_{R_\varepsilon}\varepsilon$, 
$A\approx_{R}\calT^\calG_R(A)$, and $AC\approx_R C$ in the cases
specified in Def.~\ref{def:constrans}. Since in these cases we have
 $A\sim_{R_\varepsilon}\varepsilon$, $A\sim_{R}\calT^\calG_R(A)$,
	and $AC\sim_R C$ (as follows from~(\ref{eq:simwithnf})),
	it suffices to show that
$\alpha\sim_R\beta$ implies 
$\alpha\approx_R\beta$.

So let us assume 
$\alpha\sim_R\beta$ (hence
$\calT^\calG_R(\alpha)=\calT^\calG_R(\beta)$ by~(\ref{eq:simisequiv})),
and suppose $\alpha\strans{a}_R\alpha'$; we will be done if we show
that $\beta\strans{a}_R\alpha'$.
If $a=\tau$ and $\alpha'=\calT^\calG_R(\alpha)$, then indeed
$\beta\strans{a}_R\alpha'$ since it is
$\beta\strans{\tau}_R\calT^\calG_R(\beta)$ in this case.
Otherwise we have 
$\alpha=\alpha_0\trans{\tau}\alpha_1\cdots\trans{\tau}\alpha_k\trans{a}\alpha''$
where 
$\calT^\calG_R(\alpha)=\calT^\calG_R(\beta)=
\calT^\calG_R(\alpha_0)=\cdots=\calT^\calG_R(\alpha_k)$
and $\alpha'=\calT^\calG_R(\alpha'')$.
By~(\ref{eq:simisequiv}) we have 
$\alpha\sim_R\alpha_0\sim_R\cdots\sim_R\alpha_k$, hence also 
$\alpha_k\sim_R\beta$.
Since
$\alpha_k\trans{a}\alpha''$, we must have 
$\beta=\beta_0\trans{\tau}\beta_1\cdots\trans{\tau}\beta_{k'}\trans{a}\beta''$
where 
$\beta\sim_R\beta_0\sim_R\cdots\sim_R\beta_{k'}$ and
$\beta''\sim_R\alpha''$. By~(\ref{eq:simisequiv}) we thus have 
$\calT^\calG_R(\beta)=\calT^\calG_R(\beta_0)=\cdots=\calT^\calG_R(\beta_{k'})$
and $\calT^\calG_R(\beta'')=\alpha'$; hence $\beta\strans{a}_R\alpha'$.

\bigskip

\item[(3)]%3. 
The number of states of $\calT^\calG=(Q,\var,\Delta,q_0)$
is bounded by the
number of subsets of $\var$ (hence $|Q|\leq 2^{|\var|}$).
Function $\Delta$ can be presented by $|Q|\cdot |\var|$ expressions
$\Delta(R,A)=(R',\alpha)$ where $\alpha$ (i.e., $\calT^\calG_R(A)$) is
an $R$-normal form satisfying $A\sim_R\alpha$, and thus also
$\bnorm{A}_R=\bnorm{\alpha}_R$. It is straightforward to note
that 
 $|\alpha|\leq\bnorm{\alpha}_R=\bnorm{A}_R\leq\norm{A}$, 
and $\norm{A}$ is 
at most exponential in the size of $\calG$ (by
Prop.~\ref{prop:standnorm}(5)).
The overall size of $\calT^\calG$ is thus indeed at
most exponential in the size of $\calG$.\qed
\end{enumerate}

\section{Additional remarks}\label{sec:complexity}

The main result of the paper is captured by
Theorem~\ref{th:canontrans}. 
Together with
Lemma~\ref{lem:constrans} it places the nBPA-bbis problem in
\NEXPTIME, as was discussed in Section~\ref{sec:overview}.

In the arxiv-version of~\cite{DBLP:conf/lics/CzerwinskiJ15} we
(Czerwi\'{n}ski and Jan\v{c}ar)
mentioned that a natural way for a further research is 
to look for a deterministic exponential 
algorithm that would compute the decompositions
(or a \emph{base} in the terminology
of~\cite{DBLP:conf/lics/CzerwinskiJ15}) 
by
proceeding via a certain series of decreasing over-approximations. 
In the transducer framework, this suggests to build the canonical transducer
$\calT^\calG$ by a series of stepwise refined over-approximations.

We mentioned in Section~\ref{sec:proof} that the relative equivalences $\sim_R$
(defined via the LTSs $\calL_{\calG,R}$) make sense also for general
$R\subseteq\var$, not only for $R_\gamma$, so we could think of
constructing such a more general transducer; its (exponentially many)
control states are thus given. It is then natural to
use \emph{nondeterministic} transducers $\calT$ as the over-approximations of 
$\calT^\calG$, and to try to find a method of
some safe successive decreasing of the
nondeterminism by finding where the current $\calT$ violates the consistency
and other conditions satisfied by $\calT^\calG$.
(An example of one such condition that has not been mentioned
explicitly is that
$A\in R_\gamma$ entails $A\trans{u}\eps$ for $u\in\{\tau\}^*$.)

Here we do not pursue such a task further; it
 would be interesting to clarify 
if the approach by He and Huang~\cite{DBLP:conf/lics/HeH15} can be
seen as accomplishing it.

\subsection*{Author's acknowledgements}

I would like to thank especially to Wojciech Czerwi\'{n}ski, with whom
we performed the research reported on
in~\cite{DBLP:conf/lics/CzerwinskiJ15}. 
I felt appropriate to elaborate this new version but 
Wojciech could not participate because of other duties, and
he suggested that I do this alone.
Nevertheless he commented a preliminary version of this paper, for
which I also warmly thank him.

I also thank the 
anonymous reviewers 
for their helpful comments.

\bibliographystyle{abbrv}
\bibliography{citat}

\end{document}